\documentclass[12pt,a4paper]{article}
\pdfoutput=1
\usepackage{jheppub}
\usepackage{amsfonts}
\usepackage{bbm}
\usepackage{multirow}
\usepackage{enumitem}   
\usepackage{stmaryrd}
\usepackage{placeins}
\usepackage[none]{hyphenat}
\usepackage{makecell}
\setcellgapes{3pt}

\usepackage[dvipsnames]{xcolor}

\usepackage{tikz-cd}   
\usepackage{tikz}
\usepackage{pgfplots}
\usepgfplotslibrary{fillbetween}

\def \Z{\mathbb{Z}}
\def \ZZ{\widehat{\mathbb{Z}}}
\def \R{\mathbb{R}}
\def \C{\mathbb{C}}
\def \Q{\mathbb{Q}}

\def \CA{\mathcal{A}}

\def \CD{\mathcal{D}}

\def \CN{\mathcal{N}}

\def \pt{{\text{pt}}}
\def \lk{{\ell k}}
\def \Spin{{\mathrm{Spin}}}

\def \Tor{{\mathrm{Tor}\,}}
\def \Tr{{\mathrm{Tr}\,}}
\def \Hom{{\mathrm{Hom}}}
\def \End{{\mathrm{End}}}
\def \Aut{{\mathrm{Aut}}}
\def \Ker{{\mathrm{Ker\,}}}
\def \Coker{{\mathrm{Coker\,}}}

\def \MM{{\mathfrak{M}}}

\def \q{{\mathsf{q}}}
\def \D{{\mathsf{D}}}

\renewcommand{\hat}{\widehat}

\title{$\Q/\Z$ symmetry}

\author[1]{Pavel Putrov}

\affiliation[1]{ICTP, Strada Costiera 11, Trieste 34151, Italy}

\abstract{We summarize basic features of quantum field theories with discrete symmetry $\Q/\Z$ (possibly higher form, global or gauged). The classification of representations and anomalies is quite rich and involves the ring of profinite integers. As a main example we consider in detail 3d topological Dijkgraaf-Witten $\Q/\Z$ gauge theories. We also briefly discuss relevance for some previously considered physical systems. In particular we comment on a relation to the recently discovered non-invertible symmetry in 4d QED and the problem of categorification of Chern-Simons TQFT. 
}

\begin{document}
\tikzset{->-/.style={decoration={
  markings,
  mark=at position .5 with {\arrow{>}}},postaction={decorate}}}
\maketitle

\section{Introduction and summary}
\label{sec:intro}

Recent years showed a lot of progress towards various generalizations of the usual notion of symmetry in quantum field theory (QFT): higher-form \cite{Gaiotto:2014kfa}, non-invertible, subsystem, etc (see \cite{McGreevy:2022oyu,Cordova:2022ruw} for recent review). In this paper we focus on a particular example of an invertible, possibly higher-form, abelian symmetry, which however is not of the type ordinarily considered in physics. It is a non-finitely generated discrete group $\Q/\Z$. The abelian group $\Q/\Z$ can be naturally interpreted as the torsion subgroup of $U(1)$, that is it can be considered as an internal symmetry of rotations by rational angles. We use this example to demonstrate that in many aspects non-finitely generated discrete symmetry groups can be considered on the same level as more ordinary symmetry groups, such as Lie groups or discrete finitely generated groups (which in physics are often just finite groups). At the same time, as we will see already for this relatively simple example of an infinitely generated group, the representation theory and the anomaly classification can be quite involved compared to more usual cases. 

In particular, the Pontrygin dual group of $\Q/\Z$ is $\Hom(\Q/\Z,U(1))\cong \ZZ $ is the group of profinite integers, a certain completion of the group $\Z$. It is also not finitely generated, and moreover is uncountable. This dual group appears as a new symmetry if one gauges $\Q/\Z$. In Section \ref{sec:gauge} we will also encounter  a certain non-trivial extension of $\Q/\Z$ by $\ZZ$ as a 1-form symmetry.

Some other infinite discrete symmetry groups, such as $\Z$ or $\mathrm{SL}(2,\Z)$, have been of course previously considered in some detail the literature, see for example \cite{Bhardwaj:2020ymp,Heidenreich:2021xpr}. However in those particular cases the groups are still finitely generated.

Throughout the paper, unless explicitly mentioned otherwise, we consider $\Q/\Z$ as a topological group with discrete topology, which is different from the topology induced from the standard ``continuous'' topology on $U(1)\cong \R/\Z \supset \Q/\Z$. This is in particular important when one considers representations, that is \textit{continuous} homomorphisms $\Q/\Z\rightarrow \mathrm{GL}(V)$ for some vector space $V$. For $\Q/\Z$ with the topology induced from the standard topology on $U(1)$ the representation theory would be the same as for $U(1)$ itself. Note that in principle one could also consider $\Q/\Z$ with some other topology $T$, neither discrete, nor ``continuous''. The properties of QFTs with such a symmetry then can be in principle studied by the use of the identity homomorpshim $(\Q/\Z)_\text{discrete}\rightarrow (\Q/\Z)_{T}$ considered as a map between two different topological groups (it is  continuous for any topology $T$). This homomorphism in particular induces pullbacks between the corresponding representation rings and the groups classifying anomalies.

Many techniques that we will use rely on the realization of $\Q/\Z$ as a direct limit of cyclic groups $\Z/N\Z$, which are quite well studied in the context of symmetries in QFT. The direct limit construction of $\Q/\Z$ in a sense formalizes the notion of ``large $N$'' limit of cyclic groups $\Z/N\Z$ that can appear naturally in certain families of QFTs. To obtain the classification of the anomalies we use the properties of the ``anomaly classifying functor'' which takes the symmetry groups to the abelian groups classifying anomalies. In our specific setup one can argue that it takes the direct limit of symmetry groups to the inverse limit of the groups classifying their anomalies. One can in principle consider other complicated groups realized as direct limits of more simple ones in a similar way. In particular many of our results can be easily generalized to any torsion divisible abelian group, which is necessarily of the form $\oplus_j\;\Z/{p_j}^\infty\Z$, where $p_j$ are primes and $\Z/p^\infty \Z$ is Prüfer $p$-group. Prüfer $p$-group is the subgroup of $U(1)$ consisting of roots of unity of prime power order.  The group $\Q/\Z$ is a particular example of a divisible abelian torsion group: $\Q/\Z\cong \oplus_{\text{prime }p}\;\Z/p^\infty\Z$.

Below we present the outline of the paper and summarize the results.
 In Section \ref{sec:rep} we review the basic aspects of the representation theory of $\Q/\Z$ that will be useful later. In Section \ref{sec:anomalies} we obtain classification of anomalies of $\Q/\Z$ 0-form and 1-form symmetries in $d$-dimensional QFTs for sufficiently low $d$. We consider the cases of fermionic and bosonic theories with no time-reversal symmetry. The classification of anomalies can be understood as the classification of (spin-) invertible topological quantum field theories (TQFT), or equivalently (fermionic) symmetry protected topological (SPT) orders, with symmetry $\Q/\Z$. 
 
 We start Section \ref{sec:gauge} by analyzing general features of $\Q/\Z$ gauge theories. In Section \ref{sec:QmodZ-CS} we consider in detail topological 3d Dijkgraaf-Witten-type theory with gauge group $\Q/\Z$. We analyze the spectrum of line operators and provide an explicit expression for the partition function of the theory on a closed 3-manifold. While the spectrum of line operators turns out to be rather ``wild'' (but still have an explicit description in terms of a certain infinte abelian group), the partition function turns out to be very similar, in  a sense, to the partition function of $U(1)$ abelian Chern-Simons theory. In Section \ref{sec:gauged-fermions} we consider gauging a $\Q/\Z$ symmetry in a collection of 2d fermions and analyze cancellation conditions for gauge anomaly and mixed global-gauge anomalies.
 
 In the last Section \ref{sec:applications} we briefly discuss appearance of $\Q/\Z$ symmetry in other physical systems. Specifically in Section \ref{sec:non-inv} we consider a homomorphism (in the sense described there) from the non-invertible symmetry recently found in quantum electrodynamics (QED) to $\Q/\Z$, which allows to pull-back representaions and anomalies of $\Q/\Z$ symmetry to those of the non-invertible ones. In Section \ref{sec:Habiro} we argue that elements of the Habiro ring (which unify the partition functions of 3d $SU(2)$ Chern-Simons theory on an integer homology sphere for all levels), can be interpreted as Witten indices of a supersymmetric quantum mechanics with $\Q/\Z$ symmetry.

\section{Representations}
\label{sec:rep}

In this section we review the basic aspects of the representation theory of the group $\Q/\Z$. In physics it is required to understand possible action of the symmetry on the operators and the states of the theory.

The finite dimensional irreducible representation of an abelian group $G$ are necessarily one dimensional. If one assumes unitarity, they are classified by the Pontryagin dual group $\Hom(G,U(1))$.
For the the group $U(1)$ (with the standard continious topology) they are well known to be the group of integers: 
\begin{equation}
	\Hom(U(1),U(1))\cong \Z,
\end{equation}
physically known as "chargres". The tensor product of representations  naturally corresponds to the addition operation in $\Z$. The group $\Q/\Z$ is torsion, therefore all one-dimensional representations are automatically unitary. Assuming discrete topology on $\Q/\Z$, they are known to be classified by a certain completion of $\Z$, namely the group of profinite integers $\ZZ$, playing the role of "charges" in this case:
\begin{equation}
	\Hom(\Q/\Z,U(1))\cong \ZZ.
	\label{QmodZ-dual}
\end{equation}
The standard definition of $\ZZ$ is by considering inverse limit of the cyclic groups:
\begin{equation}
	\ZZ=\varprojlim\Z/n\Z.
\end{equation}
That is, $\ZZ$ has a property of having surjective $\mod n$ reduction homomorphisms to all the cyclic groups so that all the triangles 
\begin{equation}
\begin{tikzcd}
	\Z/n\Z \ar[rr,"\mod m"] & & \Z/m\Z \\
	& \ZZ \ar[ur,"\mod n"] \ar[ul,"\mod m"] &
	\label{ZZ-mod-triangle}
\end{tikzcd}
\end{equation}
(where $m$ divides $n$) are commutative. Moreover, an element of $\ZZ$ is completely determined by its images in $\Z/n\Z$ for all $n$. That is, a ``profinite'' integer can be understood as a collection of its values modulo $n$ for all $n$, which are all consistent with each other, as they would for an ordinary integer. From this point of view it is clear that the ordinary integers from a subgroup $\Z\subset \ZZ$. The one-dimensional representation labelled by a profinite integer\footnote{Throughout the paper we will use Latin letters with hats to denote profinite integers, that is elements in $\hat{\Z}$.} $\hat{k}$ is explicitly realized as follows:
\begin{equation}
\begin{array}{rrcl}
    \rho_{\hat{k}}: & \Q/\Z&\longrightarrow & U(1),  \\
   &  \cfrac{p}{q}& \longmapsto  & \exp{2\pi i\, \cfrac{(\hat{k}\mod q)\,p}{q}},
\end{array}
\end{equation}
where $p$ and $q$ are a pair of integers\footnote{For the sake of simplicity we often write the elements of $\Q/\Z$ as their rational number representatives, suppressing explicit $\mod 1$ reduction}. From the definition of $\hat\Z$ is follows that the result does not depend on the choice of the pair $(p,q)$ representing an element of $\Q/\Z$.  We obviously get the correspondence between the tensor product of representations and the addition on $\hat{\Z}$:
\begin{equation}
    \rho_{\hat{k}}(x)\rho_{\hat{\ell}}(x)=\rho_{\hat{k}+\hat{\ell}}(x).
\end{equation}

The cosets $\Z/n\Z$ apart from the the standard addition also have a standard multiplication operation, which makes them rings. The homomorphisms in (\ref{ZZ-mod-triangle}) can  be then considered as ring homomorphisms, making $\hat{\Z}$ into a ring with $\Z$ being a subring. Moreover, since $\rho_{\hat{k}}$ takes any element of $\Q/\Z$ to a root of unity in $U(1)$ we have  $\ZZ\cong\Hom(\Q/\Z,U(1))\cong\Hom(\Q/\Z,\Q/\Z)\equiv \End(\Q/\Z)$, so that $\ZZ$, as a ring, can be identified with the ring of endomorphisms of the abelian group $\Q/\Z$.

For an element $\alpha=(p/q\mod 1)\in \Q/\Z$ we can simply write its image under the endomorphism corresponding to $\hat{k}\in\hat{\Z}$ as $\hat{k}\alpha\in \Q/\Z$, meaning
\begin{equation}
     \hat{k}\alpha \equiv \cfrac{(\hat{k}\mod q)\,p}{q} \mod 1.
\end{equation}

In order to give an example of a profinite integer which is not an ordinary one it is useful to consider a slightly different, equivalent realization of $\ZZ$, which we will do shortly.

Before that, let us note that the Pontryagin duality (\ref{QmodZ-dual}) can be easily seen from the fact that $\Q/\Z$, the group of rotations by rational angles, can be understood as the \textit{direct} limit of the cyclic groups $\Z/n\Z$ understood as groups of rotations on multiples of $2\pi/n$: 
 \begin{equation}
	\Q/\Z=\varinjlim\Z/n\Z,
	\label{QmodZ-as-limit}
\end{equation}
with the commutative triangles 
\begin{equation}
\begin{tikzcd}
	\Z/n\Z  \ar[dr,hook] & & \Z/m\Z \ar[ll,hook] \ar[dl,hook] \\
	& \Q/\Z  &
	\label{QmodZ-triangle}
\end{tikzcd}
\end{equation}
where $m$ is again assumed to divide $n$, and all the maps are the standard inclusion maps. This is just another way to say that $\Q/\Z$ is the union of all cyclic subgroups inside $U(1)$. Applying Pontryagin duality, that is $\Hom(\,\cdot\,,U(1))$ functor, to this triangle then gives (\ref{ZZ-mod-triangle}). In general, in an arbitrary category, $\Hom(\varinjlim A_n,B)=\varprojlim \Hom(A_n,B)$ so that
\begin{equation}
    \Hom(\Q/\Z,U(1))=\varprojlim\Hom(\Z/n\Z,U(1))=\varprojlim \Z/n\Z=\ZZ.    
\end{equation}
In this way one can analyze representations in an arbitrary complex vector space $V$. We have $\Hom(\Q/\Z,\mathrm{GL}(V))=\varprojlim \Hom(\Z/n\Z,\mathrm{GL}(V))$. From this one can deduce that there are no other indecomposable representations (infinite or finite dimensional) of $\Q/\Z$ apart from the one-dimensional ones described by $\ZZ$.

Alternatively to defining $\ZZ$ as the inverse limit of \textit{all} cyclic groups, one can equivalently consider it as the inverse limit only of cyclic groups of the form $\Z/n!\Z$. From this point of view a profinite integer can be explicitly given by a possibly infinite sequence of its ``digits'' in the factorial base. That is, any $\hat{k}\in \hat{\Z}$ can be uniquely represented by a formal sum\footnote{The sum is finite for positive ordinary integers, but not the negative ones.}
\begin{equation}
    \hat{k}=a_0\cdot 0!+a_1\cdot 1!+a_2\cdot 2!+a_3\cdot 3!+\ldots
\end{equation}
where $0 \leq a_i \leq i$. The sum above, although formal, have well defined values modulo $n!$ for any $n$ (or, more generally, modulo any integer) since only the first finite number of terms can contribute. The consistency of reductions modulo different integers (i.e. the commutativity of \ref{ZZ-mod-triangle}) is automatic. A non-trivial example of a profinite integer which is not an ordinary one is
\begin{equation}
    \hat{k}=0\cdot 0!+1\cdot 1!+2\cdot 2!+3\cdot 3!+\ldots
\end{equation}
Similarly, $\Q/\Z$ itself can be described as the direct limit of cyclic groups of factorial order only, with respect to the inclusions $\Z/n!\Z\hookrightarrow \Z/n!\Z$.

Due to Chinese reminder theorem we also have an isomorphism
\begin{equation}
    \hat\Z\cong \prod_{\text{prime}\;p}\hat{\Z}_p
    \label{Zhat-p-adic}
\end{equation}
where
\begin{equation}
    \hat{\Z}_p=\varprojlim\Z/p^k\Z
\end{equation}
is the ring of $p$-adic integers\footnote{The standard notation for the $p$-adic integers in the mathematics literature is just $\Z_p$, however this is a standard notation for the cyclic group with $p$ elements in physics literature. To make the distinction more apparent we use $\hat{\Z}_p$ to denote the ring of $p$-adic integers and $\Z/n\Z$ to denote a cyclic group with $n$ elements (which is also a ring).  }, with the inverse limit above taken with respect to the standard quotient maps $\Z/p^{k+1}\Z\rightarrow \Z/p^{k}\Z$. This gives yet another useful realization of $\ZZ$. It is dual to the following realization of $\Q/\Z$, already mentioned in Section \ref{sec:intro}:
\begin{equation}
    {\Q/\Z} =\bigoplus_{\text{prime }p}\,\Z/p^\infty\Z
\end{equation}
where $\Z/p^\infty\Z$ is Pr\"ufer $p$-group, the group of elements of prime power order in $U(1)$:
\begin{equation}
    \Z/p^\infty\Z=\varprojlim \Z/p^n \Z.
\end{equation}
with respect to the standard inclusions $\Z/p^n\Z\hookrightarrow \Z/p^{n+1}\Z$.

From the isomorphism (\ref{Zhat-p-adic}) one can see that $\hat{\Z}$, unlike $\Z$ is not an integral domain. That is it has divisors of zero. For example, the product of two elements of $\hat\Z$ that are supported in two different subgroups $\hat\Z_p$ and $\hat\Z_{p'}$ for two different primes is zero. This property is important for analysis of mixed anomalies between to copies of $\Q/\Z$, which will be done later in the paper.

\section{Anomalies}
\label{sec:anomalies}

According to the modern understaning the 't Hooft-like anomalies of $d$-dimensional QFTs with a given (invertible) symmetry are in one-to-one correpondence with $(d+1)$-dimensional invertible topological quantum field theories with the same symmetry. 

Consider first a general situation where we have a $d$-dimensional QFT with an (invertible) $q$-form discrete symmetry $G$, and possibly some other fixed symmetries that may non-trivially combine with space-time symmetries (such as fermionic parity, time reversal, etc.). Then the corresponding anomalies, or equivalently $(d+1)$-dimensional invertible TQFTs, are classified by a certain abelian group, which we will denote as $\CA^{d+1}(G)$. A homomorphism $G\rightarrow G'$ between a pair of groups induces a pullback $\CA^{d+1}(G')\rightarrow \CA^{d+1}(G)$ (see \cite{Witten:1983tw,Elitzur:1984kr,Ibanez:1991hv,Ibanez:1991pr,Davighi:2020bvi,Davighi:2020uab,Grigoletto:2021zyv,Grigoletto:2021oho,Davighi:2022icj} for some uses of this property). That is $\CA^{d+1}$ can be considered as a contravariant functor from the category of groups (abelian when $q>0$) to the category of abelian groups. If the group $G$ is realized as a \textit{direct} limit of a certain system of symmetry groups $G_n$ with homomorphisms $G_m\rightarrow G_n$ for $m\leq n$, then naively one can expect $\CA^{d+1}(G)$ to be the  \textit{inverse} limit of the corresponding system $\CA^{d+1}(G_n)$ with respect to the pullback maps:
\begin{equation}
    \CA^{d+1}(G)\equiv
    \CA^{d+1}(\varinjlim G_n)\stackrel{?}{=}
    \varprojlim\CA^{d+1}(G_n).
    \label{anomaly-limits}
\end{equation}
Such a property however does not hold for an arbitrary contravariant functor and an arbitrary limit of groups. For the functor $\CA^{d+1}$ classifying anomalies, we will argue that this property indeed holds if some additional conditions are imposed. 

Let us assume that the homomorphisms $G_n\rightarrow G_m$ are injective (which is the case for  $\Q/\Z=\varinjlim \Z/n!\Z$). The classifying groups $\CA^{d+1}(G)$ are certain generalized cohomology groups of the corresponding Eilenberg-Maclane space $K(G,q)$ \cite{Kitaev,Kapustin:2014tfa,Kapustin:2014dxa,Freed:2016rqq,Gaiotto:2017zba,Yonekura:2018ufj}. In other words, the functor $\CA^{d+1}$ is the composition of the covariant functor $K(\,\cdot\,,q+1)$ functor from the category of groups to the homotopy category of CW complexes with a certain generalized cohomology functor with $d+1$ being the cohomological degree (the choice cohomology theory depends on the other fixed symmetries). One can consider the sequence $K(G_1,q+1)\subset K(G_2,q+1)\subset K(G_3,q+1)\subset\ldots $ corresponding to the inclusions $G_1\subset G_2\subset G_3\subset \ldots$\footnote{See \cite{365949} for the particular case of the limit $\Q/\Z=\varinjlim \Z/n!\Z$.}. We then have $K(G,q+1)=\varinjlim K(G_n,q+1)$ with respect to the inclusions, because in such a setup the homotopy group functor is known to commute with a direct limit with respect to inclusions of topological spaces. 

A generalized cohomology functor satisfies Milnor exact sequence \cite{milnor1962axiomatic}, which, after composing with $K(\,\cdot\,,q+1)$ becomes the following short exact sequence:
\begin{equation}
    0\longrightarrow \varprojlim\nolimits^1 \CA^{d}(G_n)
    \longrightarrow \CA^{d+1}(G)
    \longrightarrow \varprojlim \CA^{d+1}(G_n)
    \longrightarrow 0
    \label{Milnor-sequence}
\end{equation}
where $\varprojlim^1$ is the ``lim-one'' operation the definition of which we review in Appendix \ref{app:lim-one}. This gives a way to calculate $\CA^{d+1}(G)$ from the knowledge of $\CA^{d+1}(G_n)$ for all $G_n$ and the pullback maps between them. In particular, we will apply it to the case of $G=\Q/\Z=\varinjlim\Z/n!\Z$ and use the known classifications of anomalies for cyclic symmetry groups. In this case it will be easy to see that the lim-one term in the Milnor sequence (\ref{Milnor-sequence}) is actually trivial (using the Mittag-Leffler condition which we also review in Appendix \ref{app:lim-one}), so that the naive isomorpshism (\ref{anomaly-limits}) actually holds.

Consider first bosonic theories with $0$-form symmetry $\Q/\Z$. For simplicity we will assume that there is no time-reversal symmetry. In general, if there is a discrete global symmetry $G$, such invertible symmetries in the ``first approximation'' are classified by the group cohomology $H^{d+1}(G,U(1))$ (or, equivalently, singular cohomology of $BG\equiv K(G,1)$). The cohomology classes correpond to choices of Dijkraaf-Witten action \cite{Dijkgraaf:1989pz} that defines the partition function of the invertible TQFT in a non-trivial background $G$ gauge field. This classification remains complete in sufficiently small dimensions $d$. 
Although the cohomology groups of $G=\Q/\Z$ can be calculated directly (e.g. using the short exact sequence of $\Z\rightarrow\Q\rightarrow \Q/\Z$, see e.g. \cite{365907}), we will use the general method above because the calculation will be easy to generalize to the complete classification given in terms of cobordism groups. The cohomology groups of cyclic groups are well known:
\begin{equation}
    H^{d+1}(\Z/n\Z,U(1))\cong \left\{
    \begin{array}{ll}
        \Z/n\Z, & d\text{ even},  \\
        0, & d\text{ odd},
    \end{array}
    \right.
\end{equation}
where we assume $d\geq 0$. For even $d$ the generating topological action on a $(d+1)$-dimensional spacetime $X$ with the background gauge field $a\in H^1(X,\Z/n\Z)$ can be written as\footnote{Throughout the paper we use the normalization for the action $S$ such that $\exp\{i\,S\}$ is the corresponding weight appearing in the partition function.}:
\begin{equation}
    \frac{2\pi}{n}\int_X a\cup \underbrace{(\beta(a)\cup \ldots \cup \beta(a))}_{d/2\text{ times}}\mod 2\pi
    \label{Zn-CS-term-even-d}
\end{equation}
where $\beta:H^1(\,\cdot\,,\Z/n\Z)\rightarrow H^2(\,\cdot\,,\Z)$
is the Bockstein cohomological operation corresponding to the short exact sequence $\Z\stackrel{n\cdot}{\rightarrow} \Z \stackrel{\mod n}{\rightarrow} \Z/n\Z$. The cup product $\cup:H^i(\,\cdot\,,\Z/n\Z)\times H^j(\,\cdot\,,\Z)\rightarrow H^{i+j}(\,\cdot\,,\Z/n\Z)$ is defined using the standard $\Z$-module structure on $\Z/n\Z$. The pullback of the inclusion map $\Z/m\Z\rightarrow \Z/n\Z$ (for $m$ dividing $n$) is the $\mod m$ map: $a\rightarrow a\mod m$. These maps are surjective and thus satisfy Mittag-Leffler condition (see Appendix \ref{app:lim-one}), therefore the lim-one term in the Milnor sequence (\ref{Milnor-sequence}) vanishes and we have
\begin{equation}
    H^{d+1}(\Q/\Z,U(1))=
    H^{d+1}(\varinjlim\Z/n\Z,U(1))=
    \varprojlim H^{d+1}(\Z/n\Z,U(1)).
\end{equation}
From the definition of $\hat{\Z}$ it immediately follows that
\begin{equation}
    H^{d+1}(\Q/\Z,U(1))\cong \left\{
    \begin{array}{ll}
        \hat{\Z}, & d\text{ even},  \\
        0, & d\text{ odd}.
    \end{array}
    \right.
\end{equation}
For $d=0$ we recover the statement $\Hom(\Q/\Z,U(1))=\hat{\Z}$ which was discussed in detail in Section \ref{sec:rep}. The topological action corresponding to a profinite integer $\hat{k}\in\hat{\Z}$ is the following:
\begin{equation}
    2\pi\, \hat{k}\cdot \int_X a\cup \underbrace{(\beta(a)\cup \ldots \cup \beta(a))}_{d/2\text{ times}}\mod 2\pi
    \label{bosonic-action-d-dimensional}
\end{equation}
 where $a\in H^1(X,\Q/\Z)$ is the background gauge field, $\beta$ is the Bockstein cohomological operation corresponding to the short exact sequence ${\Z\rightarrow \Q \rightarrow \Q/\Z}$, the cup product $\cup:H^i(\,\cdot\,,\Q/\Z)\times H^j(\,\cdot\,,\Z)\rightarrow H^{i+j}(\,\cdot\,,\Q/\Z)$ is defined using the standard $\Z$-module structure on $\Q/\Z$.

Consider now a more complete classification of anomalies of bosonic QFTs given in terms of Anderson dual to the oriented bordism groups \cite{Kapustin:2014tfa,Freed:1991bn}:
\begin{equation}
    \CA^{d+1}_\mathrm{SO}(G)\cong \Hom(\Omega_{d+2}^\mathrm{SO}(BG),\Z)\oplus
    \Hom(\Tor\Omega_{d+1}^\mathrm{SO}(BG),U(1))
\end{equation}
where the first summand is responsible for perturbative anomalies. The classification for the cyclic groups is known in low dimensions (see e.g. \cite{Kapustin:2014tfa,Hsieh:2018ifc,Garcia-Etxebarria:2018ajm,Guo:2018vij,Wan:2018bns}, note that away from 2-torsion the classification coincides with the fermionic case):
\begin{equation}
\CA^{d+1}_\mathrm{SO}(\Z/{n}\Z)=\left\{
\begin{array}{ll}
    \Z/{n}\Z, & d=0, \\
    0, & d=1, \\
    \Z/n\Z \times \Z, & d=2, \\
    0, & d=3, \\
    \Z/2\Z\times \Z/{3n}\Z\times \Z/{(n/3)}\Z, & d=4,\;3|n, \\
    \Z/2\Z\times \Z/{n}\Z\times \Z/{n}\Z, & d=4,\;3\nmid n. \\
\end{array}
\right.
\label{Zn-SO-anomaly-classification}
\end{equation}
For dimension $d< 4$ the classification (and the corresponding topological actions) essentially coincides with the classification given by the cohomology groups. The only addition is an extra $\Z$ factor in $d=2$. It corresponds to the perturbative gravitational anomaly with the degree 4 anomaly polynomial being an integer multiple of $p_1/3$, where $p_1$ is the Pontryagin class of the tangent bundle of the spacetime.

For $d=4$, the $\Z/2\Z$ factor, which is present for any $n$, corresponds to a mod 2 purely gravitational anomaly governed by the action $\pi \int w_2w_3$, where $w_i\in H^i(X,\Z/2\Z)$ are Stiefel-Whitney characteristic classes of the tangent bundle. In the case when $n$ is not a multiple of three, the classification of the remaining, not purely gravitational, anomalies is $\Z/n\Z\times \Z/n\Z$ where the generator of one of the $\Z/n\Z$ corresponds to the same topological actions as given by the group cohomology (\ref{Zn-CS-term-even-d}):
\begin{equation}
    \frac{2\pi}{n}\int_{X} a\cup \beta(a)\cup \beta(a),
    \label{Zn-CS-action-d-4}
\end{equation}
while the generator of the other $\Z/n\Z$ corresponds to action 
\begin{equation}
    \frac{2\pi}{n}\int_{X} a\cup p_1.
    \label{Zn-mixed-action-d-4}
\end{equation}

However, if $n$ is a multiple of $3$, this description breaks down\footnote{There is the following simple indication that the  actions (\ref{Zn-CS-action-d-4})-(\ref{Zn-mixed-action-d-4}) are no longer independent when 3 divides $n$. If $\beta(a)=0$ then (\ref{Zn-CS-action-d-4}) is trivial, but this also means that $a$ is a mod 3 reduction of an integer first cohomology class. Therefore it has a Poincar\'e dual represented by a smooth 4-manifold $Z$ inside a 5-dimensional spacetime $X$. Then the action (\ref{Zn-mixed-action-d-4}) can be evaluated as $\frac{2\pi}{n}\,3\sigma(Z)$, where $\sigma(Z)$ is the signature of $Z$. Therefore the second action becomes necesarily trivial if multiplied by $n/3$.} and the actual classification is given by $\Z/{3n}\Z\times \Z/{(n/3)\Z}$ with the pair of corresponding topological actions being a certain refinement of the pair of actions considered above.

Taking the inverse limit of the groups in (\ref{Zn-SO-anomaly-classification}) similarly to the case of group cohomology classification, we obtain
\begin{equation}
\CA^{d+1}_\mathrm{SO}(\Q/\Z)=\left\{
\begin{array}{ll}
    \hat{\Z}, & d=0, \\
    0, & d=1, \\
    \hat{\Z} \times \Z, & d=2, \\
    0, & d=3, \\
    \hat{\Z}\times \hat{\Z}, & d=4. \\
\end{array}
\right.
\label{QmodZ-SO-anomaly-classification}
\end{equation}
For $d\leq 3$ the topological actions are again the same as for the group cohomlogy classification, with the addition of the perturbative gravitational anomaly for $d=2$ corresponding to the degree 4 anomaly polynomial $p_1/3$. 

For $d=4$, the actions are classified by a pair of profinite integers $\hat{k},\hat{\ell}\in\hat\Z$. When $a\in H^1(X,\Q/\Z)$ does not contain 3-torsion (meaning the minimal $N\in\Z_+$ such that $Na=0$ is not divisible by 3), they can be written as
\begin{equation}
    2\pi\, \hat{k}\cdot \int_X a\cup (\beta(a)\cup\beta(a))+
    2\pi\, \hat{\ell}\cdot \int_X a\cup p_1(TX).
    \label{top-action-d-4-no-2-3-torsion}
\end{equation}

Now let us turn to the case of fermionic theories. The oriented bordism groups need to be replaced with spin-bordism groups \cite{Kapustin:2014dxa,Freed:2016rqq}:
\begin{equation}
    \CA^{d+1}_\mathrm{Spin}(G)\cong \Hom(\Omega_{d+2}^\mathrm{Spin}(BG),\Z)\oplus
    \Hom(\Tor\Omega_{d+1}^\mathrm{Spin}(BG),U(1))
\end{equation}
For cyclic groups have (see  \cite{milnor1963spin,kirby_taylor_1991,yu1995connective,bruner2010connective,Kapustin:2014dxa,Freed:2016rqq,brumfiel2016pontrjagin,Wang:2017moj,beaudry2018guide,Hsieh:2018ifc,Garcia-Etxebarria:2018ajm,brumfiel2018pontrjagin,Guo:2018vij,cheng2018classification,Wang:2018pdc,Wan:2018bns,Grigoletto:2021zyv}):
\begin{equation}
\CA^{d+1}_\mathrm{Spin}(\Z/{n}\Z)=\left\{
\begin{array}{ll}
    \Z/2\Z\times \Z/{n}\Z, & d=0, \\
    \Z/2\Z, & d=1, \\
    \Z/{2n}\Z \times \Z/2\Z\times  \Z, & d=2,\;4|n, \\
    \Z/{n}\Z \times  \Z, & d=2,\;2\nmid n, \\
    0, & d=3, \\
    \Z/{3n}\Z\times \Z/{(n/12)}\Z, & d=4,\;3|n,\;4|n, \\
    \Z/{n}\Z\times \Z/{(n/4)\Z}, & d=4,\;3\nmid n,\;4|n, \\
    \Z/{3n}\Z\times \Z/{(n/3)}\Z, & d=4,\;3|n,\;2\nmid n, \\
    \Z/{n}\Z\times \Z/{n}\Z, & d=4,\;3\nmid n,\;2\nmid n. \\
\end{array}
\right.
\label{Zn-Spin-anomaly-classification}
\end{equation}
The corresponding topological actions, especially in dimension $d+1=3$, were studied in some detail in \cite{Witten:1985xe,kirby_taylor_1991,Kapustin:2014tfa,Witten:2015aba,Guo:2018vij,Witten:2019bou,Grigoletto:2021zyv}. Taking the inverse limit gives us
\begin{equation}
\CA^{d+1}_\mathrm{Spin}(\Q/\Z)=\left\{
\begin{array}{ll}
    \Z/2\Z\times \hat{\Z}, & d=0, \\
    \Z/2\Z, & d=1, \\
    \hat{\Z} \times \Z, & d=2, \\
    0, & d=3, \\
    \hat{\Z}\times \hat{\Z}, & d=4. \\
\end{array}
\right.
\label{QmodZ-Spin-anomaly-classification}
\end{equation}
Note that $\Z/2\Z$ factor for $d=2$ does not survive in the limit, because the pullback of the inclusion $\Z/m\Z\rightarrow \Z/n\Z$ of symmetry groups (if $m|n$ and $4|m$), which can be immediately deduced from the results of \cite{Grigoletto:2021zyv}, is trivial on that factor:
\begin{equation}
    \begin{array}{rcl}
        \Z/{2n}\Z\times \Z/2\Z & \longrightarrow  & \Z/{2m}\Z\times \Z/2\Z,\\
        (a,b) & \longmapsto  & (a\mod 2m,0). \\
    \end{array}
\end{equation}
Although the maps in the inverse system are no longer surjective it is easy to see that the Mittag-Leffler condition is still satisfied. The topological actions classified by the $\hat{\Z}$ factor for $d=2$ are
\begin{equation}
    2\pi\hat{k}\, q_s(a)\mod 2\pi, \qquad \hat{k}\in \hat{\Z},
    \label{3d-fermion-anomaly}
\end{equation}
where
\begin{equation}
    q_s:\;H^1(Y,\Q/\Z)\longrightarrow \Q/\Z
    \label{quadratic-ref-coh}
\end{equation}
is a quadratic refinement of the bilinear pairing
\begin{equation}
\begin{array}{rcl}
    H^1(Y,\Q/\Z)\otimes_{\Z} H^1(Y,\Q/\Z) & \longrightarrow & \Q/\Z,\\
    a\otimes b & \longmapsto & \int_{X}a\cup \beta(b),
\end{array}    
\end{equation}
and depends on a choice of the spin-structure $s\in \Spin(X)$. Namely, it satisfies
\begin{equation}
    q_s(a+b)-q_s(a)-q_s(b)=\int_{X}a\cup \beta(b)\mod 1
\end{equation}
and in particular
\begin{equation}
    2q_s(a)=\int_{X}a\cup \beta(a)\mod 1
\end{equation}
where the right-hand side, up to the factor of $2\pi$, is the generating topological action in the bosonic case. The statement that the generating (over $\hat{\Z}$) topological action is given by $2\pi q_s$ immediately follows from the explicit universal description of the topological actions for all cyclic groups $\Z/n\Z$ considered in \cite{Grigoletto:2021zyv}. There the action was constructing by composing the embedding map $\Z/n\Z\subset \Q/\Z$ with the operation that can be identified with $q_s$. 

More concretely, consider the series of homomorphisms
\begin{equation}
H^1(X,\Q/\Z)\cong \Hom(H_1(X),\Q/\Z)\rightarrow \Hom(\Tor H_1(X),\Q/\Z)
\cong \Tor H_1(X),  
\label{QmodZ-coh-to-tor}
\end{equation}
where the second homomorpshism is dual to the inclusion $\Tor H_1(X)\subset H_1(X)$ and the last isomorphism is given by the non-degenerate linking pairing defined on closed 3-manifolds:
\begin{equation}
\begin{array}{rrcl}
    \lk: & \Tor H_1(X)\otimes \Tor H_1(X) & 
    \longrightarrow & \Q/\Z, \\
    & [a]\otimes[b] & \longmapsto & \cfrac{(a\cap c)}{p} \mod 1,
\end{array}
    \label{lk-3-manifold}
\end{equation}
where $p$ is the order of $[b]$ and $c$ is a 2-chain such that $\partial c=pb$.

The function $q_s$ is then given by the composition of the map $H^1(X,\Q/\Z)\rightarrow \Tor H_1(Y)$ given by (\ref{QmodZ-coh-to-tor}) with the quadratic refinement of (\ref{lk-3-manifold}) which was defined in \cite{kirby_taylor_1991}. 

The $\Z$ factor in the classification for $d=2$ corresponds to the standard perturbative gravitational anomaly with the degree 4 anomaly polynomial being an integer multiple of $p_1/48$.

For $d=4$, when the background $a\in H^1(X,\Q/\Z)$ does not contain 2- or 3-torsion the actions classified by $\hat{\Z}\times \hat{\Z}$ can be written in the same form as in the bosonic case, see  (\ref{top-action-d-4-no-2-3-torsion}). Otherwise one has to use certain refinements of the pair of topological invariants that appear there.

Finally, let us briefly discuss possible anomalies of $\Q/\Z$ 1-form symmetry.
We can again use the fact that $\Q/\Z=\varinjlim\Z/n\Z$ and the known results about classification of anomalies of $\Z/n\Z$ 1-form symmetries to obtain the classification of anomalies of $\Q/\Z$ as an inverse limit. In the bosonic case (as before, without time-reversal symmetry), we have
\begin{equation}
    \CA^{d+1}_{\text{1-form},\mathrm{SO}}(G)\cong \Hom(\Omega_{d+2}^\mathrm{SO}(K(G,2)),\Z)\oplus
    \Hom(\Tor\Omega_{d+1}^\mathrm{SO}(K(G,2)),U(1)).
\end{equation}
For cyclic groups the classification is the following \cite{Gaiotto:2014kfa,Gaiotto:2017yup,Wan:2018bns}:
\begin{equation}
\CA^{d+1}_{\text{1-form},\mathrm{SO}}(\Z/{n}\Z)=\left\{
\begin{array}{ll}
    \Z/n\Z, & d=1, \\
    \Z, & d=2, \\
    \Z/{2n}\Z, & d=3,\;2|n, \\
    \Z/{n}\Z, & d=3,\;2\nmid n, \\
    (\Z/2\Z)^2, & d=4,\;2|n, \\
    \Z/2\Z, & d=4,\;2\nmid n. \\
\end{array}
\right.
\label{Zn-1form-SO-anomaly-classification}
\end{equation}
The topological actions that correspond to not purely-gravitational anomalies (which are the same as for the case of ordinary symmetry discussed above) are the following:
\begin{equation}
\begin{array}{cc}
    2\pi \int_X b & (\Z/n\Z,\;d=1), \\
    \frac{2\pi}{n} \int_X b\cup b, & (\Z/n\Z,\;d=3,\;2\nmid n), \\
    \frac{2\pi}{2n} \int_X \mathcal{P}(b), & (\Z/2n\Z,\;d=3,\;2|n), \\
    {\pi} \int_X \mathrm{Sq}^2\beta_{(2,n)}b, & (\Z/2\Z,\;d=4,\;2|n), \\
\end{array}
\end{equation}
where $b\in H^2(X,\Z/n\Z)$ is the background gauge field, $\mathcal{P}:H^2(\,\cdot\,,\Z/n\Z)\rightarrow H^4(\,\cdot\,,\Z/2n\Z)$ in the Pontryagin square operation, $\beta_{(2,n)}:H^i(\,\cdot \,,\Z/n\Z)\rightarrow H^{i+1}(\,\cdot\,,\Z/2\Z)$ corresponding to the extension $\Z/2\Z\rightarrow \Z/2n\Z\rightarrow \Z/n\Z$ of the coefficients, and $\mathrm{Sq}^j:H^i(\,\cdot \,,\Z/2\Z)\rightarrow H^{i+j}(\,\cdot\,,\Z/2\Z)$ are Steenrod squares. From the explicit forms of the actions above one can conclude that the pullbacks of the inclusion map $\Z/n\Z\rightarrow \Z/nr\Z$ are the following. While for $d=1$ we have the usual mod $n$ map:
\begin{equation}
    \begin{array}{rcl}
        \Z/nr\Z & \longrightarrow & \Z/n\Z,  \\
        k & \longmapsto & k\mod n,
    \end{array}
\end{equation}
for $d=3$ it is 
\begin{equation}
    \begin{array}{rcl}
        \Z/nr\Z & \longrightarrow & \Z/n\Z,  \\
        k & \longmapsto & rk\mod n,
    \end{array}
\end{equation}
for odd $n$ and 
\begin{equation}
    \begin{array}{rcl}
        \Z/2nr\Z & \longrightarrow & \Z/2n\Z,  \\
        k & \longmapsto & rk\mod 2n,
    \end{array}
\end{equation}
for even $n$. The Mittag-Leffler condition is again satisfied\footnote{It is especially easy to see this using the prime decomposition. It is enough to consider sequences $\ldots \rightarrow \Z/p^{r+1}\Z \rightarrow \Z/p^r\Z\rightarrow \ldots $ with the map being multiplication by $p$. The images of the compositions $\Z/p^{r+k}\Z\rightarrow \Z/p^r\Z$ then stabilize to zero for $k\geq r$. }. And for $d=4$, even $n$ we have
\begin{equation}
    \begin{array}{rcl}
        \Z/2\Z & \longrightarrow & \Z/2\Z,  \\
        k & \longmapsto & 0 .
    \end{array}
\end{equation}
Unlike in the case of 0-form $\Q/\Z$ symmetry, in the inverse limit we only get a non-trivial (not purely gravitational) for $d=1$:
\begin{equation}
\CA^{d+1}_{\text{1-form},\mathrm{SO}}(\Q/\Z)=\left\{
\begin{array}{ll}
    \hat\Z, & d=1, \\
    \Z, & d=2, \\
    0, & d=3, \\
    \Z/2\Z, & d=4. \\
\end{array}
\right.
\end{equation}
The corresponding topological actions are simply
\begin{equation}
    2\pi \hat{k}\int_X b,\qquad \hat{k}\in \hat\Z
\end{equation}
where $b\in H^2(X,\Q/\Z)$.

\section{Gauging {$\Q/\Z$} symmetry}
\label{sec:gauge}

In the beginning of this section we consider some general aspects of gauging $\Q/\Z$ symmetry. First we should acknowledge that due to infinite discrete nature of the group one can expect that some of the observables (such as partition functions) in the gauged theories will be ill defined (in particular, infinite) or trivially zero, at least if defined in a naive way. However such type of problems are not unusual in QFTs. We will see that after a renormalization (if it is actually required) we can recover non-trivial quantities.

Consider first a 0-form $\Q/\Z$ gauge theory in $d$ spacetime dimensions. In the absence of charged matter it has $\Q/\Z$ electric 1-form symmetry that acts on Wilson lines which are classified by the 1-dimensional representations, that is $\hat{\Z}\cong \Hom(\Q/\Z,U(1))$. The correposnding charge operators are realized by $(d-2)$ dimensional disorder operators. Since the gauge group is discrete, the Wilson lines itself are topological and play the role of charges of $\hat{\Z}$ magnetic $(d-2)$-form symmetry. In general one can add topological theta-like terms in the action, which were considered in the Section \ref{sec:anomalies}. As we will see in the example of 3d gauge theory in Section \ref{sec:QmodZ-CS}, the presence of such terms in general can modify the symmetry structure from the direct product of electric and magnetic symmetries to their non-trivial extension.

Assuming the cancellation of anomalies (classification of which studied in Section \ref{sec:anomalies}), one can add charged matter, with fundamental fields having certain $\hat{\Z}$ valued charges. A presence of a matter field with a charge $\hat{q}\in\hat{\Z}$ will brake the electric one-form symmetry to the subgroup $\Ker\hat{q}\subset \Q/\Z$, where $\hat{q}$ is considered as a map $\Q/\Z\rightarrow \Q/\Z$, that is an element of $\End(\Q/\Z)\cong \ZZ$. If $\hat{q}=q\in \Z\subset \hat{\Z}$ this subgroup is simply $\Z/q\Z\subset \Q/\Z$. However in general this subgroup can be infinite (for example if $\hat{q}\in \ZZ_p\subset \ZZ$). Also $\Ker 
\hat{q} =0$ not only for $\hat{q}=1\in \Z\subset \ZZ$, but for any ring unit of $\hat{\Z}$, that is an element invertible under multiplication. Such units form a multiplicative group $\ZZ^\ast \subset \ZZ$ which can be naturally identified with the group of automorphisms of $\Q/\Z$ inside the ring of endomorphisms: $\ZZ^\ast=\Aut(\Q/\Z)\subset \End(\Q/\Z)=\ZZ$.

More generally one can consider $p$-form $\Q/\Z$ gauge theory. It will have $(p+1)$-form electric symmetry $\Q/\Z$ and $(d-2-p)$-form magnetic symmetry $\hat{\Z}$.

\subsection{Dijkgraaf-Witten 3d topological $\Q/\Z$  gauge theory}
\label{sec:QmodZ-CS}

Consider now in more detail a pure 0-form $\Q/\Z$ gauge theory in 3 dimensions. For simplicity, we focus on the bosonic case and we will briefly mention the modifications required for generalization to the fermionic case. The general bosonic action is of the type (\ref{bosonic-action-d-dimensional}), that is:
\begin{equation}
    S[a]=2\pi\hat{k}\int_{Y}a\cup \beta(a)\mod 2\pi,\qquad
    \hat{k}\in\hat{\Z}
    \label{QmodZ-CS-3d-action}
\end{equation}
where $a\in H^1(Y,\Q/\Z)$ is the gauge field on a 3-dimensional space-time manifold $Y$. The dynamics of the theory can be analyzed by considering it as the limit of $\Z/n\Z$ Dijkgraaf-Witten gauge theories \cite{Dijkgraaf:1989pz} with the action 
\begin{equation}
    S_n[a_n]=\frac{2\pi k_n}{n}\int_{Y}a_n\cup \beta(a_n)\mod 2\pi,\qquad
    {k}\in{\Z/n\Z},
\end{equation}
as it was done for the anomalies in Section \ref{sec:anomalies}. These theories in turn are known to be equivalent to $U(1)^2$ Chern-Simons theories with the action
\begin{equation}
    S[A]=\frac{1}{\pi}\int_Y \sum_{ij}K_{ij}A_idA_j
\end{equation}
and the level matrix\footnote{The minus sign in front of $k_n$ is due to $i$ present in the embedding map $\Z/n\Z\rightarrow U(1)$, $a\mapsto \exp\{2\pi ia/n\}$. }
\begin{equation}
    K=\left(
    \begin{array}{cc}
        -2k_n & n \\
        n & 0
    \end{array}
    \right).
\end{equation}
The eqivalence $k_n\sim k_n+n$ is realized by the field transformation $A_2\rightarrow A_2+A_1$. The fact that the integral quadratic form defined by $K$ is even corresponds to the fact that the theory is bosonic, in the fermionic case one has to replace $2k_n$ with $k_n$ defined modulo $2n$. 

It has $\det K=n^2$ independent line operators with the invertible fusion described by the group 
\begin{equation}
    \Z^2/K\Z^2\cong \cfrac{(\Z/n\Z)\times\Z}{\Z(-2k_n,n)}. 
    \label{Zn-line-group}
\end{equation}
In terms of the original $\Z/n\Z$ gauge theory, this group can be understood as an extension of the naive group  $\Z/n\Z$ of 't Hooft lines by the group $\Hom(\Z/n\Z,U(1))\cong \Z/n\Z$ of Wilson lines:
\begin{equation}
    \begin{tikzcd}
         0\ar[r] & {\Z/n\Z}\ar[r] & \cfrac{(\Z/n\Z)\times\Z}{\Z(-2k_n,n)}\ar[r]  &
        \Z/n\Z \ar[r] & 0, \\
        & a\ar[r,maps to] & {[(a,0)]}, \\
       & &   {[(a,b)]}\ar[r,maps to] & b\mod n. \\
    \end{tikzcd}
\end{equation}
In the $\Q/\Z$ gauge theory the group $\Gamma$ of line operators should then be also an extension of the naive $\Q/\Z$ group of 't Hooft lines by the group $\Hom(\Q/\Z)=\hat{\Z}$ of Wilson lines. The extension can be understood as the limit\footnote{More systematically, one can consider the more general family of extensions of $\Z/n'\Z$ by $\Z/n\Z$ realized by groups $(\Z/n\Z)\times \Z/(-2k_n,n')\Z$ (with obvious inclusion and projection maps) and take their direct limit with respect to $n'$ and inverse limit with respect to $n$. Different extensions in the system form commutative diagrams with respect to the inclusion maps $\Z/m\Z\rightarrow \Z/n\Z$ and mod maps  $\Z/n'\Z\rightarrow \Z/m'\Z$, i.e. the extensions are related by the corresponding pullbacks and pushouts respectively.} 
of the extensions (\ref{Zn-line-group}) where $k_n=\hat{k}\mod n$:  
\begin{equation}
    \begin{tikzcd}
        0\ar[r] & \hat{\Z}\ar[r] & \Gamma=
        \cfrac{\hat{\Z}\times \Q}{\Z(-2\hat{k},1)}\ar[r] &
        \Q/\Z \ar[r] & 0, \\
                & \hat{n}\ar[r,maps to] & {[(\hat{n},0)]}, \\
       & &   {[(\hat{n},\alpha)]}\ar[r,maps to] & \alpha\mod 1. \\
    \end{tikzcd}
    \label{Gamma-extension}
\end{equation}

The expectation value of a system of line operators labelled by $[(\hat{n}_i,\alpha_i)]\in \Gamma,
\;i=1,\ldots, N$ and supported on loops $\gamma_i\subset \R^3$ can be calculated again considering the family of $\Z/n\Z$ gauge theories (or, equivalently, $U(1)^2$) with couplings $k_n=\hat{k}\mod n$ for all $n$:
\begin{equation}
    \left\langle W_{[(n_1,\alpha_1)]}(\gamma_1) \cdots 
    W_{[(n_N,\alpha_N)]}(\gamma_2)\right\rangle
    =\exp 2\pi i \sum_{i,j}\lk(\gamma_i,\gamma_j)\;(\hat{n}_i\alpha_j+\hat{k}\alpha_i\alpha_j)
    \label{QmodZ-CS-loop-vev}
\end{equation}
where $\lk(\gamma_i,\gamma_j)$ is the linking number between the $i$-th and $j$-th loops. One can see that the right hand side is indeed invariant under the changes $(\hat{n}_i,\alpha_i)\rightarrow {(\hat{n}_i-2\hat{k},\alpha_i+1)}$.

One can also derive this result directly by calculating the partition function on $\R^3$ with the corresponding insertions. The Wilson line operator supported on a loop $\gamma$ and labelled by $\hat{n}\in\hat{\Z}=\Hom(\Q/\Z,U(1))$ is simply given by 
\begin{equation}
    W_{[(\hat{n},0)]}=e^{2\pi i\hat{n}\int_{\gamma} a}.
\end{equation}
Naively, a 't Hooft line operator  corresponding to an elements $\alpha\in \Q/\Z$ can be defined as a disorder operator that creates a puncture in the spacetime at its support and imposes the condition that the holonomy of the gauge field around it is $\alpha$. However the action (\ref{QmodZ-CS-3d-action}) becomes ambiguous on manifolds with boundary, because $a\cup \beta(a)$, as defined, takes values in cohomology \textit{unrelative} to the boundary, and its integration over $Y$ is then ill defined. The ambiguity can be fixed by choosing a lift of $\alpha$ to $\Q$. Let us elaborate on this. 

It is more intuitive to work Poincar\'e-Lefschetz dual homology classes. On a closed 3-manifold $Y$, without no 't Hooft operators inserted, the Poincar\'e dual of $a\in H^1(Y,\Q/\Z)$ is realized by a 2-cycle with coefficients in $\Q/\Z$. It can be always lifted to a 2-chain with coefficients in $\Q$. This 2-chain is not in general a cycle and has a non-trivial 1-chain boundary which has coefficients necessarily in $\Z\subset \Q$. This however is not in general a boundary of a 2-chain with coefficients in $\Z$, so it defines a non-trivial class in $H_1(Y,\Z)\cong H^2(Y,\Z)$ which is exactly $\beta(a)$. The result can be argued to be independent of all the choices made, in particular of the lift of the 2-cycle with coefficients in $\Q/\Z$ to the 2-chain with coefficients in $\Q$. The action $\int_Y a\cup \beta(a)$ is then computed by summing the contributions (valued in $\Q/\Z$) from the intersection points of the 2-chain dual to $a$ and 1-chain dual to $\beta(a)$.

When we introduce 't Hooft loop operators labelled by $\alpha_i\in \Q$ and supported on $\gamma_i\subset Y$ this means that Poincar\'e-Lefschetz dual of $a\in H^1(Y,\Q/\Z)\cong H_2(Y,\sqcup_i \gamma_i,\Q/\Z)$ is now represented by a 2-chain that has boundary ${\sum_i (\alpha_i\mod 1)\, \gamma_i}$. We then lift it to a 2-chain with coefficients in $\Q$ in a way that its boundary contains $\gamma_i$ with coefficients $\alpha_i\in \Q$. We then \textit{define} $\int_Y a\cup \beta(a)$ as the sum of contributions from the intersection points of that 2-chain with coefficients in $\Q$ and its boundary.

Consider now the case of insertion in $\R^3$ (or equivalently $S^3$) of Wilson-'t Hooft line operators labelled by $[(\hat{n}_i,\alpha_i)]\in \Gamma,
\;i=1,\ldots, N$ and supported on loops $\gamma_i$. Since the homology of the spacetime is trivial, there will be no nontrivial sum in the partition function. The expectation value will be just determined by the evaluation of 
\begin{equation}
    \exp 2\pi i \left(\sum_{i}\hat{n}_i\int_{\gamma_i}a
    +\hat{k}\int_{Y}a\cup \beta(a)
    \right)
\end{equation}
 in the background $a$ determined by the conditions given by 't Hooft operators. Applying the above prescription to compute the action, we recover the formula (\ref{QmodZ-CS-loop-vev}). 

The group $\Gamma$ in (\ref{Gamma-extension}) is the global 1-form symmetry group of the theory. As in the case of ordinary abelian 3d TQFTs (where line operators forn a finite abelian group) it has 't Hooft anomaly described by the bilinear pairing that appears in the expectation value (\ref{QmodZ-CS-loop-vev}) of the linked loop operators:
\begin{equation}
\begin{array}{rcl}
         \Gamma\otimes \Gamma 
& \longrightarrow & \Q/\Z,   \\
    {[(\hat{n},\alpha)]}\otimes {[(\hat{m},\beta)]}  & \longmapsto &
    \hat{n}\beta+\hat{m}\alpha+2\hat{k}\alpha\beta.
\end{array}
\end{equation}
The topological boundary conditions of the theory correspond to the Lagrangian subgroups of $\Gamma$ with respect to this pairing \cite{Kapustin:2010hk}. For an arbitrary $\hat{k}$ there is always a Lagrangian subgroup $\hat{\Z}=\{[(\hat{n},0)]\}\subset \Gamma$ formed by the Wilson lines. When $\hat{k}=0$ and $\Gamma=\ZZ\times \Q/\Z$ one can also consider Lagrangian subgroups of the form $\hat{q}\,\ZZ\times \Ker{\hat{q}}$ for an arbitrary $\hat{q}\in\ZZ$.  In this paper we will not study the full classification for general level $\hat{k}$.

Consider now the partition function on a closed connected manifold $Y$ with no inserted line operators. At least formally, using the general formula for the parturition of DW gauge theory in the case of $\Q/\Z$ gauge group we have:
\begin{equation}
    Z_{\hat{k}}(Y)=\cfrac{1}{|\Q/\Z|}\sum_{a\in H^1(Y,\Q/\Z)}
    \exp 2\pi i \hat{k}\int_Y a\cup \beta(a)
\end{equation}
where $|\Q/\Z|=|H^0(Y,\Q/\Z)|$ is the number of elements of the $\Q/\Z$ group, which is in principle infinite. In general the number of elements of $H^1(Y,\Q/\Z)\cong \Hom(H_1(Y),\Q/\Z)$ is also infinite, however the action only depends on a projection of $a$ onto the finite group $\Hom(\Tor H_1(Y),\Q/\Z)$. As was already discussed in Section \ref{sec:anomalies}, this finite group is canonically isomorphic to $\Tor H_1(Y)$ via the non-degenerate linking pairing (\ref{lk-3-manifold}). One can then explicitly separate the finite sum and an infinite overall factor:
\begin{equation}
    Z_{\hat{k}}(Y)=|\Q/\Z|^{b_1(Y)-1}\sum_{b\in \Tor H_1(Y)}
    \exp 2\pi i \hat{k}\,\lk(b,b).
    \label{Z-bosonic-QmodZ-CS}
\end{equation}

When $Y$ is a rational homology sphere one can also remove the infinity by considering the ratio
\begin{equation}
      \cfrac{Z_{\hat{k}}(Y)}{Z_{\hat{k}}(S^3)}=\sum_{b\in \Tor H_1(Y)}
    \exp 2\pi i \hat{k}\,\lk(b,b).
\end{equation}
Equivalently, one can remove a point in $Y$ and consider it as a non-compact manifold (similarly as $\R^3$ can be obtained by removing a point from $S^3$). 

In the fermionic case the linking pairing should be replaced by its quadratic refinement which depends on the choice of spin structure $s$ on $Y$ and was defined in \cite{kirby_1989}:
\begin{equation}
    q_s: \Tor H_1(Y) \longrightarrow \Q/\Z.
\end{equation}
We denote it by the same symbol as the map (\ref{quadratic-ref-coh}) since, as it was discussed in Section \ref{sec:anomalies}, it is essentially the same map. The partition function then reads
\begin{equation}
    Z_{\hat{k}}(Y,s)=|\Q/\Z|^{b_1(Y)-1}\sum_{b\in \Tor H_1(Y)}
    \exp 2\pi i \hat{k}\,q_s(b).
        \label{Z-fermionic-QmodZ-CS}
\end{equation}

Note that, up to an overall factor depending only on $b_1(Y)$, the partition functions (\ref{Z-bosonic-QmodZ-CS}) and (\ref{Z-fermionic-QmodZ-CS}) coincide with the partitions functions of respectively bosonic and fermionic $U(1)$ level $k$ Chern-Simons theory when $\hat{k}=k\in \Z\subset \hat{\Z}$. The sums correspond to the decomposition into the contributions from different critical points -- flat connections. The usual $U(1)$ Chern-Simons theory however only makes sense when the level belongs to the ordinary integers, not their profinite completion.

\subsection{2d fermions with gauged $\Q/\Z$ symmetry}
\label{sec:gauged-fermions}

Take $N_L$ left-moving and $N_R$ right-moving free Weyl fermions in two dimensions without any mass terms. Consider action of a $\Q/\Z$ symmetry on them with the respective charges $\hat{q}_{L,i}\in \hat{\Z},\,i=1,\ldots,N_L$ and $\hat{q}_{R,j}\in \hat{\Z},\,i=1,\ldots,N_R$. Similarly to the case of $U(1)$ or $\Z/n\Z$ action, such symmetry then have anomaly of the type (\ref{3d-fermion-anomaly}) with 
\begin{equation}
    \hat{k}=\sum_{j=1}^{N_R}(\hat{q}_{R,j})^2-
    \sum_{i=1}^{N_L}(\hat{q}_{L,i})^2\qquad\in \ZZ
\end{equation}
where we use the ring structure on $\hat{\Z}$ (which is compatible with the ring structure on $\Z/n\Z$ for all $n$). The symmetry can be gauged in the two dimensional theory if $\hat{k}=0$. If the condition is not satisfied, the gauged theory can be considered as the theory living on the boundary of a 3d spacetime, with the bulk theory being the $\Q/\Z$ Dijkraaf-Witten theory with the level $\hat{k}$, which was considered in detail in the previous section. One can also consider action of another copy of $\Q/\Z$ with charges $\hat{q}_{L,i}'$ and $\hat{q}_{R,j}'$. The two $\Q/\Z$ symmetries in general will have a mixed anomaly governed by the 3d topological action
\begin{equation}
    2\pi \hat{\ell}\int_X a\cup \beta(a'),\qquad a,a'\in H^1(X,\Q/\Z)
\end{equation}
with
\begin{equation}
    \hat{\ell}=\sum_{j=1}^{N_R}\hat{q}_{R,j}\hat{q}_{R,j}'-
    \sum_{i=1}^{N_L}\hat{q}_{L,i}\hat{q}_{L,i}'\qquad\in \ZZ.
\end{equation}
The second $\Q/\Z$ symmetry will remain unbroken after gauging the first one if and only if $\hat{\ell}=0$. Note that since $\hat{\Z}$ has divisors of zero the cancellation of anomalies can happen in a rather unusual way compared to the case of $U(1)$ symmetries.

Let us illustrate this in a more specific example: a single massless Dirac fermion in 2d, that is a pair of Weyl fermions. To cancel the anomaly of the first (the one to be gauged)  $\Q/\Z$ symmetry we need
\begin{equation}
    \hat{q}_L^2=\hat{q}_R^2.
    \label{3d-Dirac-anomaly-equation-1}
\end{equation}
This equation has non-trivial solutions apart from the obvious family $\hat{q}_L=\hat{q}_R$. For example, one can take $\hat{q}_R=\hat{r}+\hat{s}$, and $\hat{q}_L=\hat{r}-\hat{s}$ for some $\hat{r},\hat{s}$ such that $\hat{r}\hat{s}=0$. 

The second $\Q/\Z$ symmetry (considered as global) remains unbroken if
\begin{equation}
    \hat{q}_L\hat{q}_L'=\hat{q}_R\hat{q}_R'.
    \label{3d-Dirac-anomaly-equation-2}
\end{equation}
Suppose we choose a solution to (\ref{3d-Dirac-anomaly-equation-1}) as in the example above with $\hat{s}=0$, but with $\hat{r}$ still being a non-trivial divisor of zero. Then (\ref{3d-Dirac-anomaly-equation-2}) has solutions of the form $(\hat{q}'_L,\hat{q}'_R)=(\hat{w},\hat{w}+\hat{u})$ for any $\hat{w}$ and $\hat{u}$ such that $\hat{r}\hat{u}=0$. Of course some of the global $\Q/\Z$ transformation can be absorbed by gauge ones. In particular if one replaces $\hat{w}$ with $\hat{w}+\hat{r}$ the action will be the same.

For a general solution of (\ref{3d-Dirac-anomaly-equation-1}) one can formulate the result about unbroken symmetry that acts faithfully on the gauged Dirac fermion in the following algebraic way. Consdier the following sequence of homomorphisms:
\begin{equation}
    \begin{tikzcd}
        0\ar[r]& \Q/\Z \ar[r,"f"] &
        \Q/\Z\times \Q/\Z\ar[r,"g"]& \Q/\Z \ar[r] & 0,
        \\
        & \alpha \ar[r,maps to] & 
        (\hat{q}_L\alpha,\hat{q}_R\alpha), & & \\
        & & (\beta,\gamma) \ar[r,maps to] &
        \hat{q}_L\beta-\hat{q}_R\gamma . &
    \end{tikzcd}
\end{equation}
The group in the middle has meaning of the product of two $\Q/\Z$ groups that act separately on left- and right-moving fermions with charge one. The group $\Q/\Z$ on the left is the one that is gauged, and the map $f$ describes how it acts on the fermions. The anomaly cancellation is equivalent to the condition that the sequence above forms a complex, that is $g\circ f=0$. The unbroken part of the $\Q/\Z\times \Q/\Z$ is then given by $\Ker g$. The symmetry that acts faithfully is its quotient modulo gauge transformations, that is $\Ker g/\mathrm{Im}\, f$, the cohomology of the complex in the middle degree.

\section{Appearance in other setups}
\label{sec:applications}

This section has a more speculative nature. We consider some other types of physical systems that have $\Q/\Z$ symmetry or a closely related structure, without going too much into details. We leave more systematic and deep analysis of possible applications for the future. 

\subsection{Relation to a non-invertible symmetry in 4d QED}
\label{sec:non-inv}

In \cite{Choi:2022jqy,Cordova:2022ieu} the authors constructed an infinite family of codimension-one topological defects in a 4d QED and that act on the Dirac fermion by the axial rotation on a rational angle. Under the fusion the family forms part of a non-invertible symmetry. One can consider the following large family of codimension-1 topological operators which will be closed under the fusion\footnote{Alternatively, one can start just with the set of operators $\{\CD_\alpha,\CD_\alpha^\dagger\}_{\alpha \in \Q/\Z}$ constructed in \cite{Choi:2022jqy,Cordova:2022ieu} and then consider its closure under fusion. The new operators that will arise will be of such form.}:
\begin{equation}
    \CD_{(\D,\q;\kappa)}(M):= \exp\{2\pi i\q(\kappa)\int_M \ast j\}\,\cdot \,Z_{\D,\q;\kappa}[M;A]
    \label{QED-defect}
\end{equation}
where $M$ is the 3-dimensional submanifold supporting the operator, $j$ is the bulk axial current, $Z_{\D,\q;\kappa}$ is the partition function of an abelian 3d TQFT labelled specified by the pair $(D,\q)$, which is coupled to the bulk gauge field $A$ in a certain way specified by $\kappa$. The pair is the invariant data of an abelian TQFT which is classically realized by a Chern-Simons theory specified by a lattice $\Lambda$ \cite{Belov:2005ze,Kapustin:2010hk}. One the quantum level the theory depends on the signature of the lattice, which we assume to be zero\footnote{Starting from an arbitrary abelian TQFT one can always achieve this condition by stacking it with a multiple of $U(1)_{\pm 1}$ Chern-Simons theories. These theories are invertible if considered as spin-TQFTs. If the ambient 4d spacetime manifold has a chosen spin-structure, it induces a canonical spin-structure on the codimension-1 submanifold $M$.} in order to have vanishing framing anomaly\footnote{Generically there is no canonically induced framing on the submanifold $M$, so in order for the defect to be well defined we require this anomaly to vanish.}, the discriminant group $\D=\Lambda^*/\Lambda$, and $\q:\D\rightarrow \Q/\Z$, the quadratic refinement of the bilinear form $\D\otimes \D\rightarrow \Q/\Z$ induced from the integral bilinear pairing on the lattice. The group $\D$ plays role of the 1-form symmetry of the 3d TQFT and $\q$ specifies its 't Hooft anomaly. The extra label is the choice $\kappa \in \D$. It specifies the coupling to the bulk gauge field as follows\footnote{The theory depends only on $\kappa$ rather than its lift to $\Lambda^*$ only up to a gauge invariant local counterterm in the action, of the form $\frac{r}{4\pi}\int_Y A\wedge dA$ for an \textit{integer} $r\in \Z$}. Let $N$ be the order of the element $\kappa$ in $\D$. This means that $\kappa$ generates a subgroup $\Z/N\Z$ inside $\D$. We then identify $([dA|_{M}]/2\pi \mod N)\in H^2(M,\Z_N)$ with the background of the corresponding 1-form symmetry of the 3d TQFT. 

As was already pointed out in \cite{Belov:2005ze} the abelian 3d TQFTs form an abelian monoid under stacking. This monoid structure can be extended to the TQFTs with the specified choice of a cyclic 1-form symmetry subgroup (that is a choice of the element in the discriminant group). Namely, one can consider the abelian monoid of quadruples
\begin{equation}
    \MM=\{(\D,\q;\kappa)\}
\end{equation}
with the ``$+$'' operation defined as follows:
\begin{equation}
    (\D_1,\q_1;\kappa_1)+(\D_2,\q_2;\kappa_2):=(\D_1\oplus \D_2,\q_1+\q_2;\kappa_1\oplus\kappa_2)
\end{equation}
where $(\q_1+\q_2)(a\oplus b):=\q_1(a)+\q_2(b)$. The monoid structure is then consistent with the fusion of the topological defects (\ref{QED-defect}):
\begin{equation}
     \CD_{(\D_1,\q_1;\kappa_1)}(M) \CD_{(\D_2,\q_2;\kappa_2)}(M)=
      \CD_{(\D_1,\q_1;\kappa_1)+(\D_2,\q_2;\kappa_2)}(M).
\end{equation}
We can then consider the ``forgetful'' map
\begin{equation}
    \begin{array}{rcl}
        \MM & \longrightarrow & \Q/\Z,  \\
        \\
        (\D,\q;\kappa) & \longmapsto & \q(\kappa),
    \end{array}
    \label{monoid-to-QmodZ}
\end{equation}
which respect the addition operation, i.e. it is a homomorphism of monoids, if we forget the minus operation on $\Q/\Z$. The action of the non-invertible symmetry in QED  on the Dirac fermion of QED factors through this map.

In general a non-invertible symmetry whose fusion is described by the monoid $\MM$ is closely related to the $\Q/\Z$. In particular the representations and the anomalies of $\Q/\Z$ can be pulled back to the representations and the anomalies of $\MM$. For example, one can introduce new fields to the QED that will transform non-trivially under $\MM$ through the map (\ref{monoid-to-QmodZ}), that is their transformations will be governed by elements of $\Hom(\Q/\Z,U(1))\cong \hat{\Z}$. Moreover, if one introduces a Weyl fermion with a charge $\hat{q}\in\Z$ it will contribute to the $\Q/\Z$ 't Hooft anomaly of the type described in (\ref{QmodZ-Spin-anomaly-classification}) for $d=4$, with the corresponsing elements of $\hat{\Z}\times \hat{\Z}$ given by certain linear combinations of $\hat{q}^3$ and $\hat{q}$.

\subsection{Large color/flavor limit}

As was discussed in Section \ref{sec:rep}, the symmetry $\Q/\Z$ can be understood in a certain sense (i.e. as a direct limit) as the limit of $\Z/N\Z$ symmetries for large $N$. Such limit can naturally occur in gauge theories considered in large color or flavor limits. Such limits are in particular relevant for the analysis of holographic correspondence to the gravity.   

Consider for example an $SU(N_c)$ gauge theory in $d$-dimensions, with matter fields only in adjoint representations. It has $\Z/N_c\Z$ 1-form symmetry. So in the large color limit $N_c\rightarrow \infty$ one may consider the theory having $\Q/\Z$ 1-form symmetry. Similarly one may consider that $PSU(N_c)$ gauge theories in the large $N_c$ limit have the dual $(d-3)$-form $\hat{\Z}$  symmetry.

Alternatively one can consider $N_f$ Dirac fermions in even-dimensional spacetime, all transforming in the same representaion of some continuous gauge group $G$. The $U(1)$ axial 0-form symmetry will then be generically broken down to $\Z/{N_f}\Z$ subgroup (here we do not consider possible additional surviving non-invetible symmetries). Therefore in the large $N_f$ limit one may consider the theory having $\Q/\Z$ symmetry.

\subsection{Habiro ring and categorification of Chern-Simons}
\label{sec:Habiro}

Consider the partition function of $SU(2)$ level $k\in\Z$ Chern-Simons theory on a closed 3-manifold $Y$. It defines a topological invariant valued in $\C$, known as Witten-Reshetikhin-Turaev \cite{Witten:1988hf,reshetikhin1991invariants} invariant in mathematical literature. In the mathematical construction there is a natural generalization of this invariant $\tau_\xi(Y)$ labeled by a general root of unity $\xi=e^{\frac{2\pi ir}{k}}$, where $r$ is coprime with $k$. The invariant which corresponds to the Chern-Simons partition function is recovered when $r=1$. In principle the invariants for different $r$ are related by a Galois transformation that permutes the roots of unity. Moreover, there is a full 3d TQFT structure for an arbitrary root of unity $\xi$, realized in terms of a modular tensor category that is related to the representation theory of the quantum group (corresponding to $\mathfrak{sl}_2$ algebra) at the root of unity.  

At least formally, one can interpret the whole family of such 3d TQFTs as a single 3d TQFT with a non-trivial background $\Q/\Z$ $(-1)$-form symmetry described by the choice of\footnote{When we treat $\xi$ as an element of $\Q/\Z$ we implicitly using the isomorphism $\Q/\Z\cong \Tor U(1)$ provided by the exponential map.} $\xi\in H^0(Y,\Q/\Z)\cong \Q/\Z$ (assuming $Y$ is connected). Then one can ask a question: does this $(-1)$-form symmetry arises as a reduction of a $0$-form $\Q/\Z$ symmetry in some 4d theory, such that the partition function of the 3d TQFT on $Y$ is realized as the partition function of that 4d theory on $Y\times S^1$, with the corresponding charge operator inserted at $Y\times \pt$ where $\pt$ is a point on $S^1$? Since we do not want to have dependence on the size of $S^1$ we will assume that the effective quantum mechanics obtained by the reduction of the 4d theory on $Y$ is supersymmetric and $S^1$ has periodic boundary conditions on the fermions. Then the statement above is equivalent to the statement that the 3d partition function function is realized as a trace over the Hilbert space of the 4d TQFT on $Y$ of the quantum operator representing the symmetry transformation by $\xi\in \Q/\Z$.  This question is very closely related to the problem of finding a categorification of the WRT invariant (which would be in a sense similar to the Khovanov's categorification \cite{khovanov2000categorification} of Jones polynomial), and essentially one of the ways to formulate it.

If such a statement was true, in particular it would imply that $\tau_\xi(Y)$, as a function of $\xi\in \Q/\Z$ is a difference of characters of complex representations (in general reducible) of $\Q/\Z$. That is it is an element of representation ring $R(\Q/\Z)$ over $\C$. 

When $Y$ is an integer homology sphere, as we will see below, such statement is indeed true but with a modification: $\tau_\xi(Y)$ is an element of the completion $\hat{R(\Q/\Z)}:=\varprojlim  R(\Z/n\Z)$. In the completion we consider the inverse limit of the system representation rings of cyclic groups with respect to the pullbacks of the inclusions $\Z/m\Z\rightarrow \Z/n\Z$ (where, as often in this paper, $m$ is assumed to divide $n$). Note that the representation ring functor does not commute with the limits: 
\begin{equation}
    \hat{R(\Q/\Z)}:=\varprojlim  R(\Z/n\Z)\neq R(\varinjlim\Z/n\Z)=R(\Q/\Z).
\end{equation}
However, for any complex, possibly infinite dimensional vector space $V$ we have $\Hom(\Q/\Z,GL(V))=\varprojlim \Hom(\Z/n\Z,GL(V))$. This means that the completion $\hat{R(\Q/\Z)}$ does not actually introduce any new representations of $\Q/\Z$, however it essentially allows to have differences of infinite dimensional representations such that when restricted to any $\Z/n\Z$ subgroup the difference simplify to the difference of finite-dimensional representations. Physically this means that fermions and bosons in the Hilbert space form infinite dimensional representations of $\Q/\Z$, however when the action of the symmetry is restricted to any finite cyclic subgroup, all but the finite number of bosons and fermions can be combined into pairs of the same representation and lifted from the zero energy level. The ring $R(\Q/\Z)\cong \Z[\ZZ]$ is a subring in the completion $\hat{R(\Q/\Z)}$. An example of its element in the completion which is not an element of $R(\Q/\Z)$ can be given by the formal infinite sum $\sum_{m\geq 1}([0]-[m!])$ where $[\hat{m}]$ denotes the one-dimensional representation labelled by charge $\hat{m}\in \ZZ$. Its reduction to any $R(\Z/n\Z)$ is well defined and given by the finite sum $\sum_{m=1}^{n-1}([0]-[m!])$. Using that $R(\Z/n\Z)\cong \Z[q]/(1-q^n)$ we can write explicitly 
\begin{equation}
    \hat{R(\Q/\Z)}=\varprojlim \Z[q]/(1-q^n).    
\end{equation}
 That is one can understand it as a certain completion of the ring of polynomials in a formal variable $q$.

In \cite{habiro2008unified} Habiro defined a topological invariant $\hat{\tau}$ of an integer homology spheres\footnote{The analog for rational homology spheres was constructed in \cite{beliakova2011unified}
.} which is valued in a different completion of the polynomial ring:
\begin{equation}
    \hat{\tau}(Y)\in \hat{\Z[q]}:=\varprojlim  \frac{\Z[q]}{(1-q)(1-q^2)\cdots (1-q^n)}.
\end{equation}
Its elements can be written as formal infinite series of the form
\begin{equation}
    \hat{f}=\sum_{m\geq 0}f_m(q)(1-q)(1-q^2)\ldots (1-q^m)
\end{equation}
where $f_m(q)\in \Z[q]$ are polynomials (not unique for a given element in $\hat{\Z[q]}$). There is a well defined evaluation at roots of unity map:
\begin{equation}
    \begin{array}{c}
        (\;\cdot\;)|_\xi:\ \hat{\Z[q]}\;\longrightarrow  \C, \\
         \hat{f}|_\xi\;:=\;\sum_{m= 0}^{k-1} f_m(\xi)(1-\xi)(1-\xi^2)\ldots (1-\xi^m),
    \end{array}
\end{equation}
where $k$ is the order of $\xi$. The Habiro invariant unifies WRT invariants for all roots of unity:
\begin{equation}
    \hat{\tau}(Y)|_{\xi}=\tau_{\xi}(Y).
\end{equation}
Moreover, by their definition, the evaluation maps factor through the ring homomorphisms:
\begin{equation}
    \begin{array}{rcl}
        \hat{\Z[q]}& \longrightarrow  & \Z[q]/(1-q^k) \\
         \hat{f}& \longmapsto & \sum_{m= 0}^{k-1} f_m(q)(1-q)(1-q^2)\ldots (1-q^m) \mod (1-q^k).
    \end{array}
\end{equation}
These homomorphisms commute with the maps in the system defining the inverse limit $\hat{R(\Q/\Z)}=\varprojlim  \Z[q]/(1-q^n)$. This implies that they define a unique homomorphism
\begin{equation}
    \hat{\Z[q]}\;\longrightarrow \; \hat{R(\Q/\Z)}.
\end{equation}
Moreover, since the values of the evaluation maps are known to completely fix the element of the Habiro ring \cite{habiro2004cyclotomic} 
, this homomorphism must be injective, so $\hat{\Z[q]}$ can be understood as a subring of $\hat{R(\Q/\Z)}$. This means that there is indeed an element of the completed representation ring $\hat{R(\Q/\Z)}$ (provided by the image of $\hat{\tau}$) the evaluations of which at roots of unity recover all WRT invariants.

Note that the fact that $\hat{\Z[q]}$ is only a part of $\hat{R(\Q/\Z)}$ is related to the fact that the elements the Habiro ring can be interpreted as ``analytic functions'' on $\Q/\Z$, in particular they have certain nice property with respect to \textit{Habiro topology} on $\Q/\Z$. 

Finally let us point out that there are physical proposals about categorification of the WRT invariants via certain supersymmetric 4d/5d  coupled system, that has a $U(1)$ symmetry that commutes with the supercharge \cite{Witten:2011zz,Gukov:2016gkn,Gukov:2017kmk}. The flavored Witten index of this system, where $q$ plays the role of $U(1)$ fagacity then is supposed to recover the WRT invariant for the root of unity $\xi=e^\frac{2\pi i}{k}$ in the limit $q\rightarrow \xi$. Namely one considers a twisted $\CN=2$ 5d $SU(2)$ super-Yang-Mills theory on $Y\times S^1\times [0,+\infty)$ with a Nahm pole boundary condition at $Y\times S^1\times \{0\}$. 

The $U(1)$ symmetry is the instanton symmetry in 5d. The precise definition of the Witten index in such system however involves various extra choices, including the choice of asymptotic condition on the 5d fields at $+\infty$. It is also not possible to completely decouple the 5d bulk degrees from the boundary preserving the $U(1)$ symmetry since there is a non-trivial inflow of ABJ-type anomaly from the bulk. It would be interesting to analyze more systematically if there is indeed some concrete physical mechanism of breaking of $U(1)$ down to $\Q/\Z$, such that the $U(1)$ flavored Witten index in the 4d/5d coupled system reduces to the element of the Habiro ring when $Y$ is an integer homology sphere\footnote{The situation is in a sense similar to the one reviewed in Section \ref{sec:non-inv}, however there is no direct analog of the construction of \cite{Choi:2022jqy,Cordova:2022ieu} when the gauge theory is non-Abelian. Naively, to construct the analog of the defect labelled by a rational number $n/k=1/(m_1-1/(m_2-\ldots )$ one could try to use the 3d $\CN=4$ topologically twisted  theory of $T^{m_1}ST^{m_2}S\ldots S$ duality wall in $\CN=4$ $SU(2)$ 4d SYM. This symmetry has $SU(2)\times SU(2)$ global symmetry (ignoring the global group structure). Then one can gauge one of the $SU(2)$ inside the defect and the other $SU(2)$ identified with the bulk gauge symmetry (i.e. the bulk gauge field is treated as background in the theory living on the defect). In the case of $n=1$ such 3d theory was considered in \cite{Creutzig:2021ext}. However the 3d theory remains to be actually topological only when the background $SU(2)$ connection is flat, which cannot be always the case, as it is identified with dynamical field in the bulk.}

\subsection{Galois symmetry}

The group $\hat{\Z}=\Hom(\Q/\Z,U(1))$, dual to $\Q/\Z$ (that is it appears if one gauges $\Q/\Z$), plays an important role in Galois theory. In particular for an algebraic closure of $\overline{\mathbb{F}_{p}}$ of a finite field $\mathbb{F}_{p}$ of prime order $p$ we have $\mathrm{Gal}(\overline{\mathbb{F}_p}/\mathbb{F}_{p})\cong \hat{\Z}$, generated by the Frobenius automorphism $\mathrm{Fr}:\,x\mapsto x^p$. If $X$ is a variety over $\mathbb{F}_p$, this Galois group acts on  $X(\overline{\mathbb{F}_{p}})$, the corresponding algebraic variety over $\overline{\mathbb{F}_{p}}$. This in particular gives a way to calculate the number of points of $X$ over $\mathbb{F}_{p^n}$ by  Lefschetz trace formula\footnote{For some recent appearance of such counts in physically relevant context see for example \cite{Candelas:2019llw,Kachru:2020sio,Bonisch:2022mgw}. As this topic is far from being the focus of this article we do not provide a complete set of references here, they can be found in the mentioned papers.}:
\begin{equation}
    \# X({\mathbb{F}_{p^n}})=\Tr_{H^*(X(\overline{\mathbb{F}_{p}}),\Q_\ell)} \,(-1)^F \,(\mathrm{Fr^*})^n
\end{equation}
where $H^*(\ldots)$ is the $\ell$-adic \'etale cohomology, $F$ is the cohomological degree modulo 2, and $\mathrm{Fr}^*$ is the induced action of the Frobenius automorphism on the cohomology. If one were to interpret this formula physically, that is as the trace over the a Hilbert space\footnote{The local zeta function, which combines the counts for all $n$, then can be understood as the trace over the Fock space built on this Hilbert space.}, with $X(\overline{\mathbb{F}_{p}})$ being an analog of the target space of supersymmetric quantum mechanics, then $\mathrm{Gal}(\overline{\mathbb{F}_p}/\mathbb{F}_{p})\cong \hat{\Z}$ would play a role of physical symmetry.

\section*{Acknowledgements} 
I would like to thank B.~Acharya, A.~Beliakova, T.~Dimofte, A.~Grigoletto, S.~Gukov, A.~Klemm, D.~Pei, F.~Rodriguez-Villegaz, I.~Runkel, S.-H.~Shao for  discussions.

\appendix

\section{Lim-one}
\label{app:lim-one}
In this section we provide the definition of $\varprojlim {}^1$ operation and its basic properties. Consider an inverse system of abelian groups:
\begin{equation}
    \ldots \stackrel{f_3}{\longrightarrow} A_3\stackrel{f_2}{\longrightarrow} A_2\stackrel{f_1}{\longrightarrow} A_1 \stackrel{f_0}{\longrightarrow} A_0.
    \label{A-inverse-system}
\end{equation}
Consider the map
\begin{equation}
    \begin{array}{rrcl}
        f:& \prod_{n\geq 0} A_n & \longrightarrow & \prod_{n\geq 0} A_n \\
        & (a_0,a_1,a_2,\ldots) & \longmapsto &
        (a_0-f_0(a_1),a_1-f_1(a_2),a_2-f_2(a_3),\ldots).
    \end{array}
\end{equation}
Then, by definition:
\begin{equation}
    \varprojlim  A_n:=\Ker f,
\end{equation}
\begin{equation}
    \varprojlim\nolimits^1 A_n:=\Coker f.
\end{equation}
The Mittag-Leffler condition can be formulated as follows. Assume that for any $n\geq 0$ there exists $k\geq n$ such that for any $m\geq k$ the image of the map $f_n\circ \ldots \circ f_{m-1}:A_m\rightarrow A_n$ is the same as the image of $f_n\circ \ldots \circ f_{k-1}:A_k\rightarrow A_n$.  Then it follows that $\varprojlim ^1 A_n=0$. The Mittag-Leffler condition means that for any $n$ the images of the maps $A_k\rightarrow A_n$ stabilize for large $k$. It is in particular satisfied if all the maps in the system (\ref{A-inverse-system}) are surjective.

\bibliographystyle{JHEP}
\bibliography{QmodZ-bib}

\end{document}